\documentclass[12pt]{iopart}

\usepackage{latexsym}
\usepackage{graphics}
\usepackage{epsfig}
\usepackage{graphicx}
\usepackage{bm}

\begin{document}
\title{Permanent magnetic microtraps for ultracold atoms}
\author{Amir Mohammadi, Saeed Ghanbari and Aref Pariz}
\address{Department of Physics, Faculty of Science, University of Zanjan, 45371-38791 Zanjan, Iran}
\eads{\mailto{amir.mohammadi@uni-ulm.de}, \mailto{sghanbari@znu.ac.ir}, \mailto{aref\_pariz@znu.ac.ir}}

\begin{abstract}
We propose and numerically study two permanent magnetic micro-structures for creating Ioffe-Pritchard microtraps.
A bias magnetic field is used to vary the depth, trap frequencies and the minimum of each microtrap.
After the Bose-Einstein condensation achievement, the bias magnetic field can be slowly removed to increase the trap
barrier heights for more efficiently holding the Bose-Einstein condensates.
Even without the external magnetic field, it is possible to hold ultracold atoms in the microtraps.  These microtraps may also be useful for single atom experiments for quantum information processing due to their very high confinement.
\end{abstract}



\section{Introduction}
After the experimental realization of the Bose-Einstein condensation (BEC) in 1995~\cite{BECAnderson1995,BECDavis1995}, the study of
ultracold Bose and Fermi gases has become one of the most active fields of research in atomic physics~\cite{Pethick02,Legget2001,Metcalf02,Barret2001,KaiSunNat2012}.
Neutral atoms can be trapped in the non-zero minimum of a magnetic field gradient. Magnetic microtraps
can produce this spatial gradient to confine and guide neutral atoms in ultracold regimes and Bose-Einstien condensates~\cite{Hinds98,Dekker2000,Nguyen2005}. Current-carrying wires are common tools to create magnetic microtraps~\cite{wein95,Ess98}. Large number of
experiments have been done by using wire traps, ranged from single microtrap to arrays of microtraps to manipulate
ultracold atoms and to realize the BEC~\cite{Nguyen2005,Fort98,Yin02,Muller99,Grabowski03,Reichel99,Du04,Fortagh07}. Beside wire traps,
 technological developments have provided us with high quality magnetic microstructures based on permanent magnetization as
an alternative approach for creating lattices of microtraps for ultracold atoms~\cite{Ricc94,Hinds99,Ghanbari06,Boyd07,Gerritsma07,Ghanbari07,Fernholz08,Schmied12}.
While wire-based traps involve high current densities, leading to trap loss and heating of atoms near surfaces,
permanent magnetic microtraps can produce highly stable potential wells with low noise. Atom chips using permanent
magnetic slabs are also preferred over wire-based magnetic traps for their better confinement features due to their high curvatures and for the freedom to vary the layout, design and magnitudes of the magnetic potentials. Like wire traps,
permanent magnetic microtraps offer arrays of periodic microtraps~\cite{Ghanbari06,Gerritsma07,Ghanbari07,Mohammadi13} and single microtrap~\cite{Fernholz08}. Many experimental works in cold
atom physics with permanent magnetic traps have been reported and BEC has been demonstrated in various setups
such as F-shaped permanent magnetic atom chip and 2D permanent magnetic lattices~\cite{Du04,Fernholz08,Gar10,Weber03,Cornell96,Hall06,vul98,brad97}.
Generally, atom chips are microscopic integrated matter-wave devices which can produce flexible potentials in space
~\cite{Ghanbari06,Hall06,Du09,Folman02,ott01,cano08}. They are used for investigation of the physics of correlated many-body quantum systems~\cite{Greiner02}, matter-wave interferometry~\cite{Fortagh07,Wang05,Jo07,Schumm05} and phase transitions~\cite{Ghanbari09,Guglielmino10}. Moreover, atom chips are suitable for the implementation of scalable quantum information processing~\cite{Cirone05,Birkl07,Tejada05}. Recently, multi-particle entanglement on an atom chip was experimentally realized, which is a useful resource for quantum metrology~\cite{Riedel10}. \\
In this paper, we propose two simple Ioffe-Pritchard permanent magnetic atom chips for holding ultracold atoms
and Bose-Einstein condensates.  The microtraps are ’self-biasing’ and external bias fields can be used to change the magnetic
field potential barrier heights for loading and manipulating ultracold atoms. Clouds of atoms can be moved
adiabatically from a quadruple magnetic trap and loaded into the microtraps by varying external magnetic fields.
This particular loading trajectory helps to avoid any secondary minimum (e.g., see ~\cite{Fernholz08}) as much as possible. The reverse
process can be done to release cloud of atoms from trap. The main motivation of this work is introducing simple
magnetic microtraps to hold atoms in ultracold regimes without bias field and offering a simple tool for changing the
height of potential barriers in the microtrap. Similar to other experiments, using Radio Frequency (RF) waves, clouds
of ultracold atoms are forced to evaporate during loading process or after that ~\cite{Pethick02,Fortagh07,Fernholz08}. However, there is also another
way for ejecting energetic atoms from trap by tuning external magnetic field. This method is not recommended
in our case due to low gain of evaporative cooling which is not in all three directions of the trap, but can be used beside other
methods for effective cooling.  We have performed numerical calculations for two relatively low and high magnetization values 800 and 3800 G. The results are presented for structures in different scales to give some perspective to experimentalists.  Our calculations for $^{87}\hspace{-0.03cm}$Rb atoms predict relatively high BEC transition temperatures (i.e. several $\mu$ K) which could be interesting. The method
has been proposed in a novel simple three-wire-based magnetic microtrap\cite{Du09}, where barrier heights of the trapping
potential are lowered so that the atoms with higher energies are removed from the trap~\cite{Pethick02,Fortagh07}.\\
\section{U- and H-like-shaped fully permanent atom chips}

\begin{figure}[b]
\begin{center}
\includegraphics[width=7 cm]{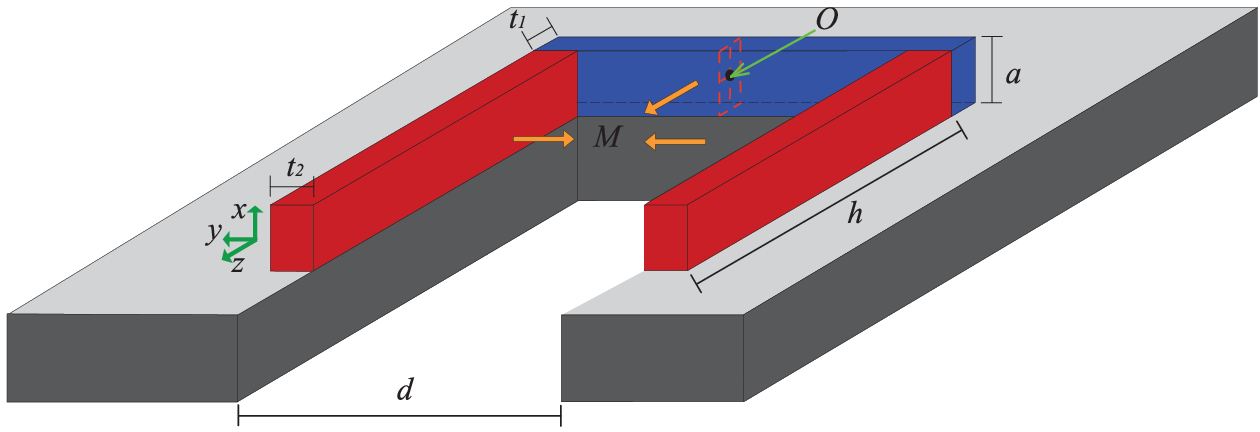}
\end{center}
\caption{
(color online). Schematic of a U-shaped microtrap. Thicknesses of the middle slab (blue) and the parallel ones (red slabs) are $t_1$ and
$t_2$, respectively. Separation between the parallel slabs is $d$. Magnetization is perpendicular to the biggest area of the magnets (orange arrows) and has same value for all
slabs. Length and width of all slabs are $h$ and $a$, respectively.}
\label{figure1}
\end{figure}
figure 1 shows a U-shaped microstructure consisting of three permanent magnetic slabs. This structure can produce Ioffe-Pritchard microtraps. The two parallel (red) slabs have opposite magnetization along $y$ direction while the middle (blue) one has same magnetization in the $z$ direction. Magnetic field due to the parallel slabs is cancelled near the
 center of the trap. By adding the third slab with magnetization in the $z$ direction, minimum of the magnetic field is increased to $B_{min}\neq0$.  Middle slab has a thickness $t_1$, the parallel ones have same thickness $t_2$ and all slabs have same length $h$ and width $a$.  Separation between the parallel slabs is $d$.  This structure can be made by depositing every single magnetic slab on a larger nonmagnetic material (gray objects) on a desired scale, as we have shown in figure 1 and figure 3.  In this paper, a perpendicular magnetization with two values $4\pi$M$\sim 800{\hspace{0.07cm}}$G and $4\pi$M$\sim 3800{\hspace{0.07cm}}$G has been assumed for all magnetic slabs.  Materials such as Alnico alloys ~\cite{Geo02}(e.g., Alnico-5) and $Tb_6$$Gd_{10}$$Fe_{80}$$Co_4$ ~\cite{Ghanbari06} with high Curie temperatures can be used for the magnetization values 800 and 3800 G, respectively.
One can deposit the magnetic material in a strong magnetic field, below Curie temperature,
to orient the magnetization in the desired direction and finally put all parts together to form the magnetic structures~\cite{mant07,nalwa02}.
\begin{figure}[ht]
\begin{center}
    \includegraphics[width=9cm]{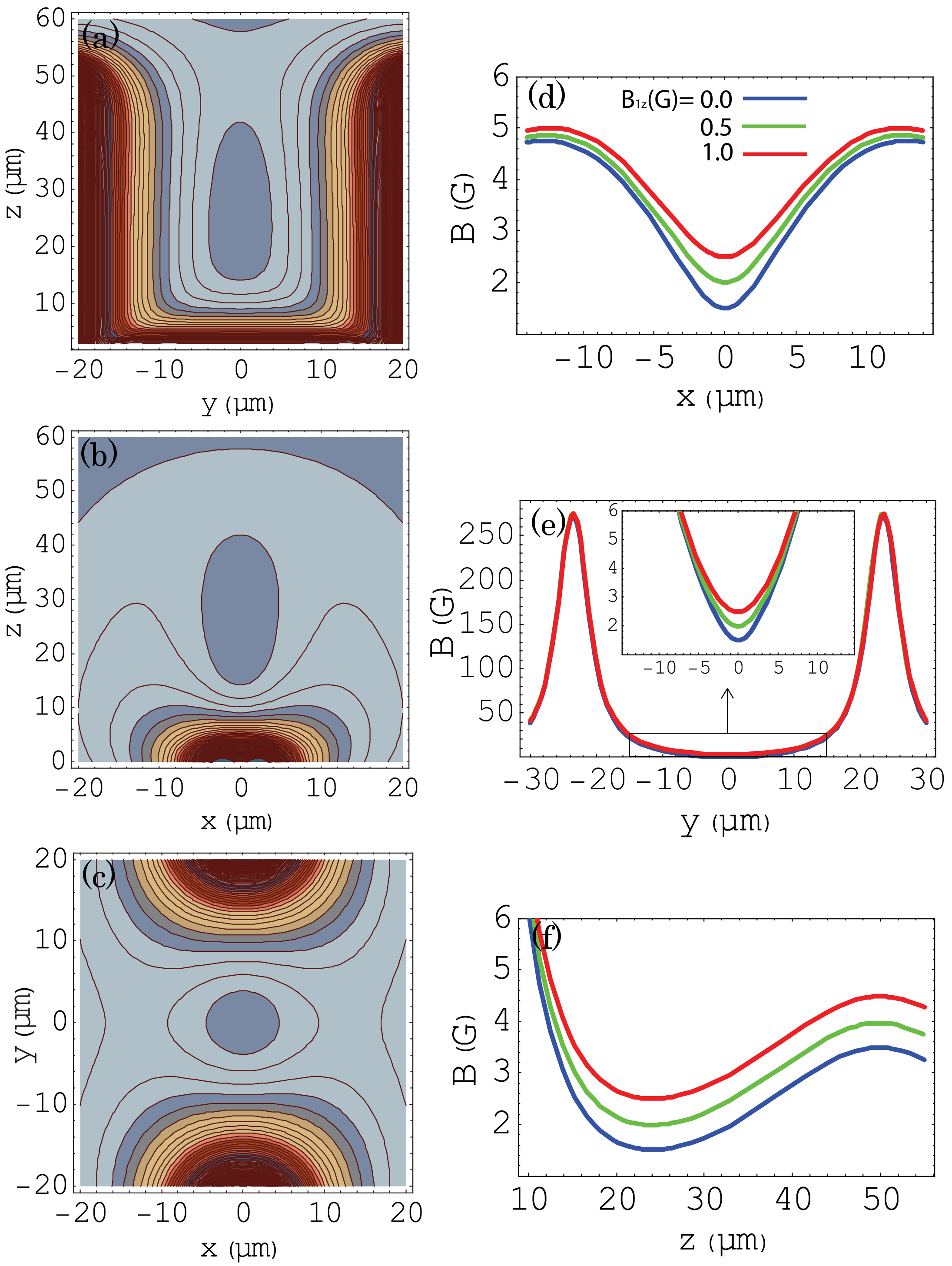}
    \caption{(color online). (a), (b) and (c) Contour plots of $B$, for the U-shaped microtrap described in figure 1, at $x=x_{min}$,$y=y_{min}$  and $z=z_{min}$
planes, respectively, when the external bias field is zero. Center of the microtrap is at $x_{min}$ =$y_{min}$ = 0 and $z_{min}$ = $24\mu$m. (d), (e) and (f) Plots
of magnetic field along $x$, $y$ and $z$ directions, respectively, for $B_{1z}$ = 0, 0.5 and 1.0 G. Plots show variation of $B$ near the center of the microtrap.
$B_{min}$ has a the nonzero value of 1.5 G even when the bias field is zero.  Parameters used for these plots are $h=50 \mu$m, $a=5 \mu$m, $d= 44 \mu$m, $t_1$ =  $2\mu$m and $t_2$ = $3 \mu$m and $4\pi$M$\sim 800{\hspace{0.07cm}}$G.}
\label{figure2}
\end{center}
\end{figure}
The interaction potential energy between the atoms and the magnetic field is given by $U={\hspace{0.07cm}}\mu_{{}_B}g_{{}_{ F}}m_{{}_F}B$ where ${\hspace{0.07cm}}\mu_{{}_B}$, $g_{{}_F}$, $m_{{}_F}$ and $B$ are the Bohr magneton, Land$\acute{\rm e}$ factor, magnetic quantum number and magnetic field modulus, respectively.
We consider $^{87}\hspace{-0.03cm}$Rb atoms when the cloud is compressed (CMOT) and optically pumped to the quantum state $F=2$ and $m_{{}_F}=2$ with the Land$\acute{\rm e}$ factor $0.5$~\cite{Pethick02}.
To do the numerical calculations, we have written a Mathematica code and have used the software package Radia~\cite{Radia} interfaced to Mathematica. The trap frequencies are determined for $^{87}\hspace{-0.03cm}$Rb atoms using the relations $\omega_{ i}=\gamma \sqrt{\partial^{2}B/\partial {x_i}^{2}}$ at the minimum of the potential where $x_1=x$, $x_2=y$, $x_3=z$, $\omega_1=\omega_x$, $\omega_2=\omega_y$, $\omega_3=\omega_z$ and $\gamma=\sqrt{{\hspace{0.07cm}}\mu_{{}_B}g_{{}_F}m_{{}_F}/m}$.  Here, $m$ is the $^{87}\hspace{-0.03cm}$Rb atom mass.
To investigate how an external magnetic field can change the trap frequencies and also depth of the trap, an external magnetic field $\bf{B}_{1}$ = $B_{1z}\hat{z}$ is applied. We start from a relatively small trap structure (Scale$A_U$) with $h=50\mu$m, $a=5\mu$m, $d=44\mu$m, $t_1= 2\mu$m, $t_2= 3\mu$m and low magnetization $4\pi$M$\sim 800{\hspace{0.07cm}}$G.
According to figure 2, we have an Ioffe-Pritchard microtrap for this configuration and an external magnetic field in the $z$
direction can change the depth of the microtrap and frequencies in the $x$ and $y$ directions but leaves its position and $z$
unchanged. Moreover, the height of the potential barrier in each direction depends on the bias magnetic field. figure 2
also shows that the U-shaped permanent magnetic atom chip can produce a single magnetic microtrap either with a bias field
or without it which may have applications in quantum information processing and data storage. For understanding the
behaviour of the trap on different scales and magnetization values, we repeat our calculations for relatively high magnetization
$4\pi$M$\sim 3800{\hspace{0.07cm}}$G on different scales. For the bias magnetic field $B_{1z}$ = 0, the trap frequencies $\omega_x$/2$\pi$, $\omega_y$/2$\pi$ and $\omega_z$/2$\pi$ for both values of magnetization are given in \tref{table1}.  This table shows that for $B_{1z}\neq0$ all the frequencies are less than the corresponding values for $B_{1z}$ = 0. Moreover, according to the \tref{table1}, second derivatives of the magnetic field modulus i.e. curvatures of $B$ along $x$ and $y$ directions, decrease during the increase of the bias field from $B_{1z}$ = 0.  \Tref{table1} shows that for equal magnetization, trap frequencies in small traps are bigger. For all scales, trap minimum depends on the magnetization of the structures and only changes when external bias field in $z$ direction is applied.

$\triangle U^{x}/k_{ B}$, $\triangle U^{y}/k_{ B}$ and  $\triangle U^{z}/k_{ B}$ for all different trap configurations with and without $B_{1z}$ are compared in \tref{table2} where $\triangle U^{i}$=$U_{ max}^{i}$-$U_{ min}^{i}$.  Here, $i$=1, 2 and 3 correspond to $x$, $y$ and $z$, respectively.  According to the table, this microtrap is highly asymmetric with respect to the barrier heights in the $x$
and $y$ directions. Energy level spacings  $\hbar\omega_{ i}/k_{ B}$ along the $x_i$ direction are other parameters which we have calculated for this microtraps in \tref{table2}. This table also clearly shows that traps with higher magnetization have larger potential
barrier heights and can be used for better holding, cooling and trapping ultracold atoms. Actually, tables~\ref{table1} and~\ref{table2}
indicate that the potential barrier heights and trap minimum almost linearly increase by increasing the magnetization.\\
Another single permanent magnetic microtrap consisting of three rectangular permanent magnetic slabs with two different thicknesses of $t_1$ and $t_2$ and separation of $d$ in a slightly different arrangement with respect to the U-shaped
configuration is also investigated. Schematic of this structure is shown in figure 3.  According to figure 4, by changing a
bias field ${\bf B_{ 1}}=B_{ 1z}\hat{z}$, it is possible to vary the barrier heights in the H-like-shaped atom chip.

\begin{table*}[t]
 \caption{\label{table1}Numerical results for U-shaped fully permanent magnetic chip (figure  1).  Results are obtained in different magnetization values for $^{87}$Rb in F=2 and $m_F$=2 state.  On scale A$_U$ we have $h=50 \mu$m, $a=5 \mu$m, $d= 44 \mu$m, $t_1$ =  $2\mu$m and $t_2$ = $3 \mu$m, while scales B$_U$ and C$_U$ are two and four times larger than scale A$_U$, respectively.}
\begin{center}
\small\addtolength{\tabcolsep}{-5pt}
\begin{tabular}{ccccc}
\hline
Parameter&Scale A$_U$&Scale A$_U$&Scale B$_U$&Scale C$_U$ \\
$B_{1z}$(G)&0\hspace{0.9cm}1\hspace{0.2cm}&0\hspace{0.9cm}3\hspace{0.2cm}&0\hspace{0.9cm}3\hspace{0.2cm}&0\hspace{0.9cm}3\hspace{0.2cm}\\
\hline
$4\pi M$(G) &800&3800&3800&3800 \\
\verb  $\frac{\partial^2 B}{\partial x^2}\hspace{.1cm}({G\over{cm}^2})$   & $2.5\times 10^5$ \hspace{0.4cm}$1.4\times 10^5$&$1.2\times 10^8$ \hspace{0.4cm}$8.1\times 10^7$& $2.9\times 10^7$ \hspace{0.4cm}$2.0\times 10^7$ &$7.4\times 10^6$ \hspace{0.4cm}$5.1\times 10^6$  \\
\verb  $\frac{\partial^2 B}{\partial y^2}\hspace{.1cm}({{\rm G}\over{{\rm cm}}^2})$ &$2.6\times 10^5$ \hspace{0.4cm}$1.6\times 10^5$&$1.2\times 10^8$ \hspace{0.4cm}$8.7\times 10^7$& $3.1\times 10^7$ \hspace{0.4cm}$2.1\times 10^7$ &$7.7\times 10^6$ \hspace{0.4cm}$5.4\times 10^6$  \\
\verb  $\frac{\partial^2 B}{\partial z^2}\hspace{.1cm}({{\rm G}\over{{\rm cm}}^2})$   & $1.6\times 10 ^{4}$& $1.6\times 10 ^{6}$&$1.8\times 10 ^{6}$&$4.6\times 10 ^{5}$ \\
\verb  $\omega_x$/2$\pi$(kHz)  & 6.4\hspace{1cm}4.9   & 1.3\hspace{1cm}1.15& 0.69\hspace{1cm}0.57 &0.34\hspace{1cm}0.28       \\
\verb  $\omega_y$/2$\pi$(kHz)  &6.5\hspace{1cm}4.5     & 1.4\hspace{1.1cm}1.2&0.71\hspace{1cm}0.59&0.35\hspace{1cm}0.29      \\
\verb  $\omega_z$/2$\pi$(kHz)  &1.6     & 0.24&0.17&0.08      \\
\verb  $ z_{min}$   ($\mu$m)   & 24& 24&72&96.2 \\
\verb  $ B_{min}$   (G)        &  1.5\hspace{1cm}2.5&7.13\hspace{1cm}10.13& 7.13\hspace{1cm}10.13&7.13\hspace{1cm}10.13\\
\hline
\end{tabular}
\end{center}
\end{table*}

\begin{table*}[t]
 \caption{\label{table2} More details about U-shaped microtraps. Values have been obtained numerically for $^{87}$Rb in F=2 and $m_F$=2 state.}
\begin{center}
\begin{tabular}{cccc}
\hline
Parameter&Definition&Scale A$_U$&Scale A$_U$ \\
$B_{1z}$(G)&z component of bias field&0\hspace{1cm}1\hspace{0.2cm}&0\hspace{1cm}3\hspace{0.2cm}\\
\hline
$4\pi M$(G)&Magnetization of slabs&800&3800 \\
 $\Delta U^x$/$k_B$ (mK)&Potential barrier height&0.22\hspace{.9cm}0.17&1\hspace{.9cm}0.87\\
 &in $x$-direction&&\\
  $\Delta U^y$/$k_B$ (mK) &Potential barrier height&19&87\\
   &in $y$-direction&&\\
  $\Delta U^z$/$k_B$ (mK) &Potential barrier height&0.13&0.67\\
   &in $z$-direction&&\\
  $\hbar \omega_x$/$k_B$ ($\mu$K)&Energy level spacing in &2.0\hspace{.5cm}1.5&0.42\hspace{.5cm}0.35 \\
  &$x$-direction &&\\
  $\hbar \omega_y$/$k_B$ ($\mu$K) &Energy level spacing in &2.0\hspace{.5cm}1.5&0.42\hspace{.5cm}0.35  \\
    &$y$-direction &&\\
  $\hbar \omega_z$/$k_B$ ($\mu$K)&Energy level spacing in&0.5&1.0  \\
    &$z$-direction &&\\
\end{tabular}
\end{center}
\end{table*}
This configuration of permanent magnetic slabs differs from the previous one in terms of fabrication and also
possibility of movement of the magnetic slabs which allows for mechanically varying (coarse adjustment) the trap depth and frequencies.
Numerical results given in figure 4 show that, this H-like-shaped microstructure can produce a minimum of magnetic
potential in the space between parallel slabs above the middle one. These slabs can be set up on a table with any
possible and desired separations and again a 3D Ioffe-Pritchard magnetic microtrap can be created even without
a bias magnetic field.
If we change $B_{1z}$, the trap frequencies change.  \Tref{table3} shows these values for the H-like-shaped atom chip
with both values of low magnetization $4\pi$M$\sim 800{\hspace{0.07cm}}$G and high magnetization $4\pi$M$\sim 3800{\hspace{0.07cm}}$G on different scales. As plots in figures 4(d), 4(e) and 4(f) show, change of the external field does not change location of the microtrap center. Curvatures
of the magnetic field close to the center of the microtrap also change by variation of the bias field, as \tref{table3} shows.
Curves in figure 4 have been plotted for this configuration on scale $A_H$ with $h=80\mu$m, $a_1=2\mu$m, $a_2=3\mu$m, $d=16\mu$m, $t_1 = 200nm$ and $t_2 = 250nm$, where
$4\pi$M$\sim 800{\hspace{0.07cm}}$G.  Initial and final values of the potential barrier heights $\triangle U^{i}/k_{ B}$ and energy level spacings $\omega_i$/2$\pi$ are also presented in \tref{table4}.
\begin{figure}[h]
\begin{center}
\includegraphics[width=7cm]{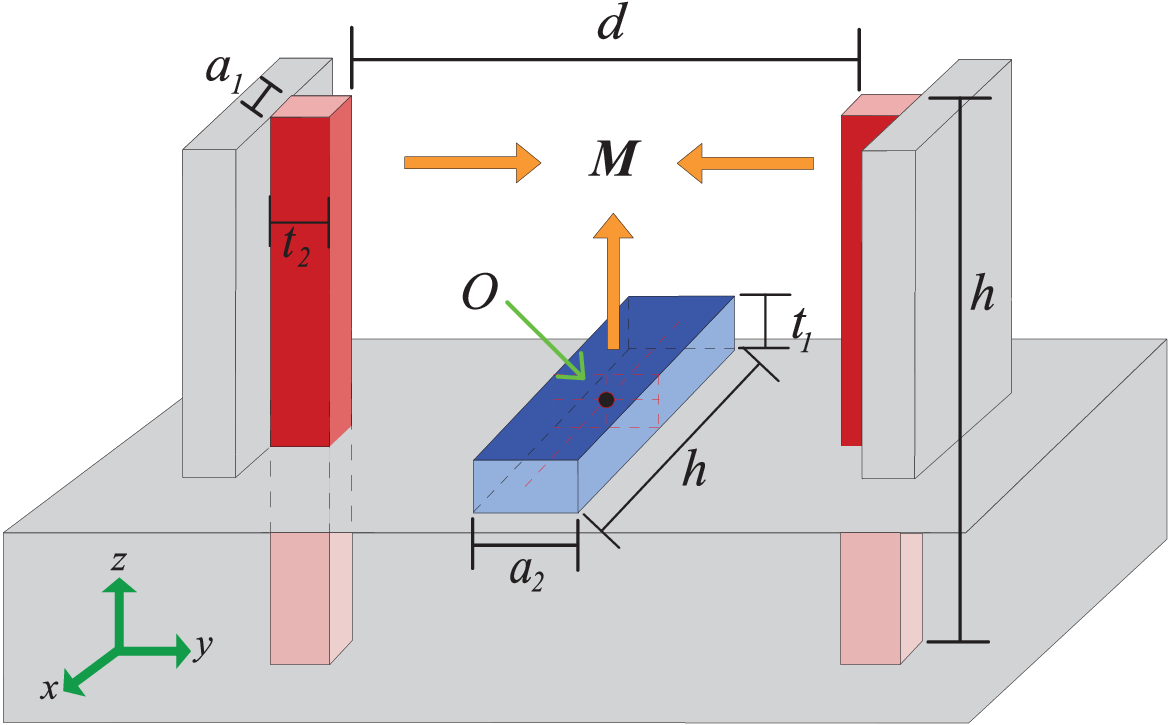}
\caption{(color online). Schematic of an H-like-shaped permanent magnetic microtrap. The middle (blue) slab has a thickness of $t_1$ and thickness
of the parallel (red) ones is $t_2$. Also, separation between the parallel slabs is $d$.  Magnetization is perpendicular to slabs (Orange arrows).  All three
slabs have same length $h$. Width of the parallel slabs is $a_1$ while that of the other slab is $a_2$.}
\label{figure3}
\end{center}
\end{figure}
Our investigations show that, movement of the parallel slabs in the H-like-shaped microstructure can also change
the depth of the microtrap. It is possible to mechanically reduce the depth of the microtrap via increasing the separation
between the parallel slabs. However it should be done adiabatically to avoid heating of atomic cloud. Previously, various methods have been used to load Bose-Einstein condensates and Fermi
gases into microtraps; loading clouds of atoms at constant phase space density by magnetic transfer is one of the
adiabatic transfer methods~\cite{Fortagh07}.  U- and H-like-shaped traps may also be useful for the displacement of clouds of atoms.  There are lots of
useful methods for the creation of these structures, like: Physical Vapour Deposition, Pulsed Laser Deposition and also
Additive Patterning Lithography.
\begin{figure}[ht]
\begin{center}
\includegraphics[width=8.5cm]{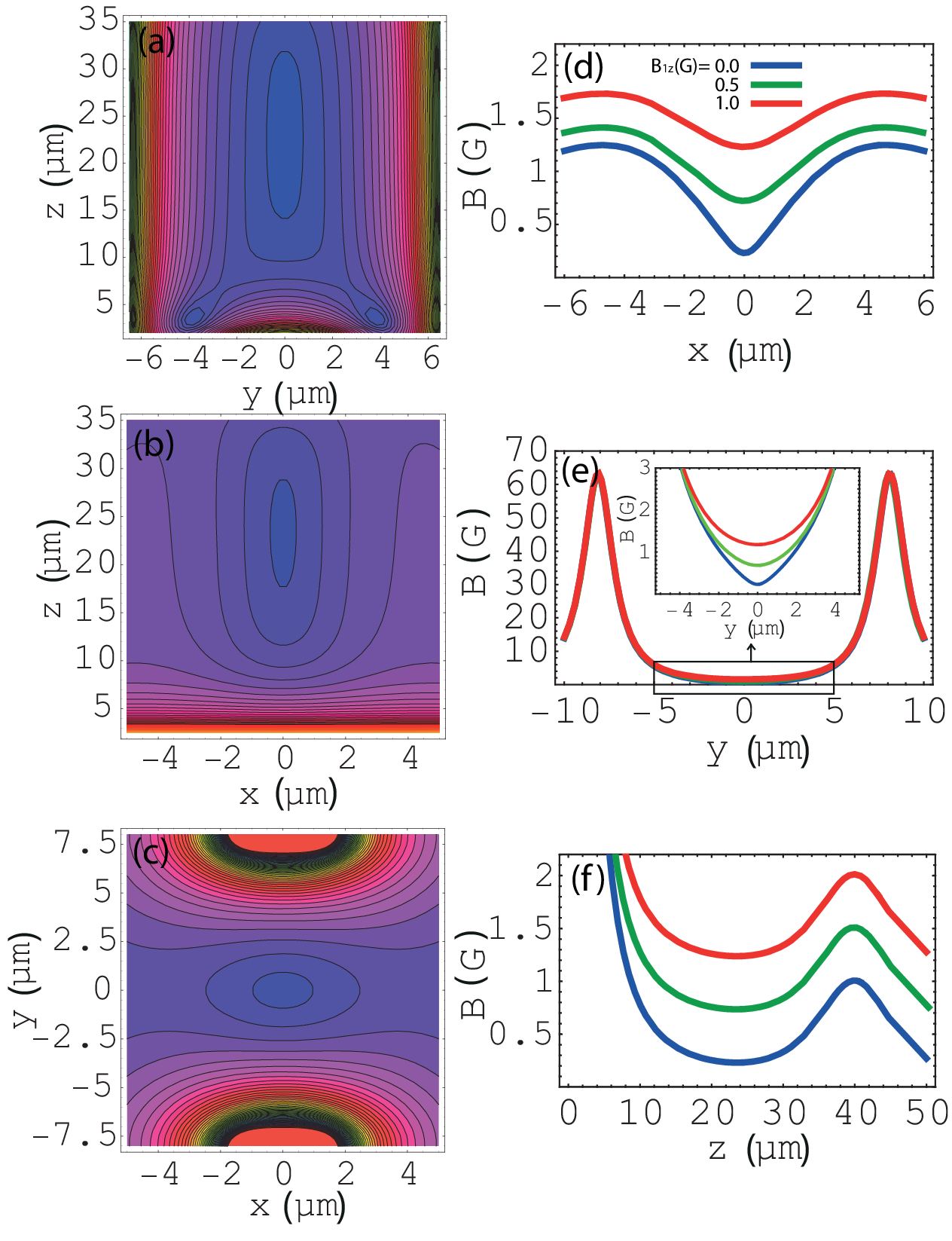}
\caption{
(Color online) (a), (b), (c) Contour plots of magnetic fields in $x=x_{ min}$, $\it y=y_{ min}$ and $\it z=z_{ min}$ planes, respectively
when external bias field is zero.  Center of the microtrap is at $x_{ min}=y_{ min}=0$ and $z_{ min}=23.6{\hspace{0.07cm}}\mu$m above the middle slab.
(d), (e) and (f) Plots of magnetic field along $x$, $y$ and $z$ directions, respectively, for $B_{1z}=0$, $0.5$ and $1.0{\hspace{0.07cm}}$G.
By increase of the bias magnetic field, the potential barrier heights in the $x$ and $y$ directions are reduced.
Moreover, the bias field does not change the position of the microtrap center.
Furthermore, $B_{ min}$ is $0.23{\hspace{0.07cm}}$G when the bias field is zero.  This figure describes the magnetic field modulus for the H-like-shaped atom chip shown in figure 3 for relatively small configuration
$h$=80$\mu$m, $a_1$=2$\mu$m, $a_2$=3$\mu$m, $d$=16$\mu$m, $t_1$ = 200nm and $t_2$ = 250nm, where $4\pi$M$\sim 800{\hspace{0.07cm}}$G.} \label{figure4}
\end{center}
\end{figure}
\section{Holding ultracold atoms and Bose-Einstein condensates}

A BEC transition temperature of $T_c=1.6{\hspace{0.1cm}}{\hspace{0.07cm}}\mu$K at a trap depth of $11 {\hspace{0.1cm}}{\hspace{0.07cm}}\mu$K
via RF evaporative cooling has been reported for $^{87}\hspace{-0.03cm}$Rb atoms in an F-shaped fully permanent magnetic atom chip~\cite{Fernholz08}.
Without considering the interactions between atoms, for a 3D anisotropic harmonic oscillator, $T_c$ is given by~\cite{Pethick02}
\begin{equation}\label{equation1}
T_{c}\approx 4.5(\frac{\overline{f}}{100Hz})N^{1/3} nK,
\end{equation}%
where $\overline{f}$ and $N$ are the geometric mean of the three trap frequencies $f_i=\omega_i/2\pi$, and number of atoms, respectively.  In the presence of interactions, we should include the correction
\begin{equation}\label{equation2}
{{\delta T_{c}}\over{T_{c}}}\approx-0.73{{\omega_{ m}}\over{\overline{\omega}}} N^{-1/3}-1.33{{a}\over{\overline{a}}}N^{1/6},
\end{equation}
where $\overline{a}$, $\omega_{ m}$ and $a_s$ are $\sqrt{\hbar/2\pi m\overline{f} }$, ($\omega_{ x}+\omega_{ y}+\omega_{ z})/3$ and s-wave scattering length, respectively~\cite{Giorgini96}.
Since $\delta T_{c}/T_{c}$ has a negative value, the corrected transition temperature is less than the temperature given by equation (1). In the absence of any bias field, taking into account the interactions between atoms, BEC critical temperature for $10^6$ $^{87}$Rb atoms in different configurations U- and H-like-shaped microtraps, are predicted in the ranges  6.9-11$\mu$K and 12-16$\mu$K, respectively.  These relativity high BEC transition temperatures for $10^6$ $^{87}$Rb atoms are obtained due to high curvatures in center of the microtraps,
which is essential property of permanent magnetic microtraps.  However, these temperatures compared with experimentally achieved
value of 1.6$\mu$K are not improbable.  If high magnetization is used for microtraps, then height of the magnetic potential
energy lets us to confine thermal atoms. For example, U-shaped trap with high magnetization of $4\pi$$M=3800 G$ can trap cold
atoms with mean temperature of approximately 150 $\mu$K.  This temperature is roughly doppler cooling limit of Rb atoms in MOT.  Therefore It is possible to directly load the atomic cloud from MOT to this trap and hold them without any bias field.  With RF field and increasing external magnetic field, we can perform evaporative cooling.  When the bias field is slowly increased, the barrier height in the $x$ direction is lowered and atoms in the higher energy levels leave the trap. In the U and H-like-shaped traps, increasing the external magnetic field can reduce the trap depth and decrease the average
temperature of atoms down to around 10 $\mu$K and below where numerical calculations predict that we attain BEC.
During evaporative cooling, trap requires non-zero minimum for preventing
spin flip loss.  After achieving the BEC, the bias field must be slowly removed in order to increase the trap barrier
heights for holding the Bose-Einstein condensate.

In high density ranges under p$\approx10^{-11}$ Torr, two-body spin relaxation and three-body recombination are two dominant loss mechanisms. The corresponding loss rate is obtained by ~\cite{Pethick02}.
\begin{equation}\label{equation3}
\Upsilon=\frac{1}{n}\frac{dn}{dt}\approx nG+Ln^2,
\end{equation}
where n, G and L are density, the two-body spin-relaxation and three-body-recombination coefficients, respectively.  For density of alkali atoms around n=$10^{12}-10^{14}$ atoms/cm$^3$, loss rate is usually about 10$^{-3}$-$10^{-1}$ s$^{-1}$.  As an example, on scale A for the U trap ($A_U$ in \tref{table1} ) with particle number N=$10^6$ for $^{87}$Rb atoms, density is $3\times10^{13}$ atoms/$cm^3$ and we have $\Upsilon\sim$ 0.05$s^{-1}$, while for the scale $C_U$, density reduces to $6.4\times10^{12}$ atoms/$cm^3$ and relatively low loss rate of $\Upsilon\sim7\times10^{-3}$$s^{-1}$ is obtained which allows for achieving a BEC with high atom numbers.  Here, G $<10^{-15}$cm$^3$/s ~\cite{Boesten96} and L=$1.8\times10^{-29}$cm$^6$/s~\cite{soding99}.

\begin{table*}[t]
 \caption{\label{table3}Numerical results for H-like-shaped fully permanent magnetic atom chip shown in figure 1.  Results are obtained for $^{87}$Rb in F=2
and $m_F$=2 state with two values of magnetization $4\pi$M$\sim 800{\hspace{0.07cm}}$G and $4\pi$M$\sim 3800{\hspace{0.07cm}}$G. On scale A$_H$, we have $h$=80$\mu$m, $a_1$=2$\mu$m, $a_2$=3$\mu$m, $d$=16$\mu$m, $t_1$ = 200nm and $t_2$ = 250nm. Scale B$_H$ is two times larger than scale A$_H$.}
\begin{center}
\begin{tabular}{ccccc}
\hline
Parameter&Scale A$_H$&Scale A$_H$&Scale B$_H$\\
$B_{1z}$(G)&0\hspace{1cm}1\hspace{0.2cm}&0\hspace{1cm}3\hspace{0.2cm}&0\hspace{1cm}3\hspace{0.2cm}\\
\hline
$4\pi M$(G) &800&3800&3800\\
\verb  $\frac{\partial^2 B}{\partial x^2}\hspace{.1cm}({G\over{cm}^2})$   & $8.9\times 10^5$\hspace{1cm}$1.6\times 10^5$&$4.2\times 10^6$\hspace{1cm}$1.14\times 10^6$&$1.07\times 10^6$\hspace{1cm}$2.8\times 10^5$ \\
\verb  $\frac{\partial^2 B}{\partial y^2}\hspace{.1cm}({G\over{cm}^2})$ &$8.9\times 10^5$\hspace{1cm}$1.6\times 10^5$&$4.2\times 10^6$\hspace{1cm}$1.14\times 10^6$&$1.07\times 10^6$\hspace{1cm}$2.8\times 10^5$ \\
\verb  $\frac{\partial^2 B}{\partial z^2}\hspace{.1cm}({G\over{cm}^2})$   &$3.8\times 10^3$&$4.4\times 10^3$&$3.8\times 10^3$ \\
\verb  $\omega_x$/2$\pi$(kHz)  &12\hspace{1cm}5.2 &26.3\hspace{1cm}13.6 &13\hspace{1cm}6.8    \\
\verb  $\omega_y$/2$\pi$(kHz)  &12\hspace{1cm}5.2 &26.3\hspace{1cm}13.6 &13\hspace{1cm}6.8   \\
\verb  $\omega_z$/2$\pi$(kHz)  &0.78    & 1.71 & 0.8       \\
\verb  $ z_{min}$   ($\mu$m)   &  23.6& 23.6 &47.17 \\
\verb  $ B_{min}$   (G)        &  0.23\hspace{1cm}1.23&1.1\hspace{1cm}4.1& 1.1\hspace{1cm}4.1\\

\hline
\end{tabular}
\end{center}
\end{table*}

\begin{table*}[t]
 \caption{\label{table4} More details about H-like-shaped microtraps. Values have been obtained numerically for $^{87}$Rb in F=2 and $m_F$=2 state.}
\begin{center}
\begin{tabular}{cccc}
\hline
Parameter&Definition&Scale$A_H$&Scale$A_H$ \\
$B_{1z}$(G)&z component of bias field&0\hspace{1cm}1\hspace{0.2cm}&0\hspace{1cm}3\hspace{0.2cm}\\
\hline
$4\pi M$ (G) &Magnetization of slabs&800&3800 \\
 $\Delta U^x$/$k_B$ ($\mu$K)&Potential barrier height &70\hspace{1cm}34&250\hspace{1cm}140\\
 &in $x$-direction&&\\
  $\Delta U^y$/$k_B$ (mK)&Potential barrier height &4.2&20.15\\
   &in $y$-direction &&\\
  $\Delta U^z$/$k_B$ ($\mu$K)&Potential barrier height &55&260\\
   &in $z$-direction&&\\
  $\hbar \omega_x$/$k_B$ ($\mu$K) &Energy level spacing in &0.58\hspace{1cm}0.25&1.26\hspace{1cm}0.65 \\
  &$x$-direction&&\\
  $\hbar \omega_y$/$k_B$ ($\mu$K)&Energy level spacing in &0.58\hspace{1cm}0.25&1.26\hspace{1cm}0.65\\
    &$y$-direction&&\\
  $\hbar \omega_z$/$k_B$ ($\mu$K)&Energy level spacing in&0.03&0.08  \\
    &$z$-direction&&\\
\end{tabular}
\end{center}
\end{table*}

\section{Conclusions}
To summarize, we have introduced two fully permanent magnetic atom chips to confine, hold and manipulate ultracold atoms. We have suggested using external magnetic fields together with RF waves to reduce potential barrier heights of the permanent magnetic microtraps for more efficient evaporative cooling. Considering the high magnetic field curvatures in permanent microtraps compared with those of current-carrying wires and optical traps, the BEC critical temperature $T_C$ is predicated to locate in the $\mu$K ranges however loss mechanisms are strong. Potential barrier heights of the microtraps linearly depend on the magnetization of slabs. For cooling and trapping thermal clouds of atoms, highly magnetized materials must be used in microtrap structures to increase the minimum of trap, potential height and trap frequencies. If smaller magnetic curvatures and trap frequencies are required in certain potentials heights, then large scale configurations can be useful.  These large scale atom chips provide relatively large volumes for achieving the BEC with high atom numbers and low atom loss.  Due to their very high confinement, these microtraps may be suitable for single atom experiments for quantum information processing.  Another advantage of these microtraps is that they can store ultracold atoms and Bose-Einstein condensates without any external bias field. These permanent magnetic atom chips may be used together with other atom chips as promising tools for the investigation of quantum degenerate gases.  Such applications of these new types of microtraps should have a good impact on the development of the atom chip field.  The fully permanent magnetic micro-structures presented here are very simple and the dimensions and magnetization of the magnetic slabs are within the accessible ranges of the contemporary technology. Depending on the magnetization of slabs, cold atoms can be loaded directly or after intermediate cooling stages from the MOT to these microtraps.
\section*{Acknowledgments}
The authors would like to thank Peter Hannaford, Bryan Dalton and Brenton Hall for helpful discussions and comments.\\

\end{document}